\documentclass[aps,11pt]{revtex4}
\usepackage[colorlinks=true,linkcolor=blue,urlcolor=blue,filecolor=black,citecolor=red,pdfstartview=FitV,pdftitle={},pdfsubject={},
pdfkeywords={},pdfpagemode=None,bookmarksopen=true]{hyperref}
\usepackage{graphicx}
\usepackage{epstopdf}%
\usepackage{amsmath}
\usepackage{amsfonts}
\usepackage{amssymb}
\usepackage{color}%
\usepackage{dcolumn}
\usepackage{slashed}
\usepackage{amssymb,ulem}
\usepackage{float}
\usepackage{amsthm,amsmath,amssymb}
\usepackage{mathrsfs}
\usepackage{subfigure}
\usepackage{wrapfig}
\usepackage{indentfirst}
\usepackage{multirow}
\usepackage{appendix}
\makeatletter

\newcommand{\Rmnum}[1]{\expandafter\@slowromancap\romannumeral #1@}

\makeatother

\begin{document}
\title{Thermodynamics of charged black holes in Maxwell-dilaton-massive gravity}
\author{Rui-Hong Yue$^{1}$\footnote{rhyue@yzu.edu.cn;}, Kai-Qiang Qian$^{1}$\footnote{kaiqiangqian@outlook.com}, Bo Liu$^{2,3}$\footnote{fenxiao2001@163.com;}, De-Cheng Zou$^{4,1}$\footnote{dczou@jxnu.edu.cn}}

\affiliation{$^{1}$Center for Gravitation and Cosmology, College of Physical Science and Technology, Yangzhou University, Yangzhou 225009, China\\
$^{2}$School of Physics, Northwest University, Xi'an, 710127, China\\
$^{3}$School of Physics and Information Science, Shaanxi University of Science and Technology, Xi'an, 710021, China\\
$^{4}$College of Physics and Communication Electronics, Jiangxi Normal University, Nanchang 330022, China}

\date{\today}

\begin{abstract}
\indent
Considering the nonminimal coupling of the dilaton field to the massive graviton field in Maxwell-dilaton-massive gravity, we obtain  a class of analytical solutions of charged black holes, which are neither asymptotically  flat nor (A)dS.  The  calculated thermodynamic quantities, such as mass, temperature and entropy, verify the validity of the first law of black hole thermodynamics. Moreover, we further investigate the critical behaviors of these black holes in the grand canonical and canonical ensemble, 
and find a novel critical phenomenon never before observed, known as the ``reverse" reentrant phase transition with a tricritical point. It implies that the system undergoes a novel ``SBH-LBH-SBH" phase transition process, and  is the reverse of the ``LBH-SBH-LBH" process observed in reentrant phase transitions.

\end{abstract}


\maketitle

\section{Introduction}
\label{1s}

The universe is expanding at an accelerating rate, which is in contrast with the deceleration predicted by the standard Friedmann model. This observation, based on data of supernovae  \cite{Riess:1998,Perlmutter:1999} and cosmic microwave background
(CMB) radiation \cite{Planck:2015fie,WMAP:2003elm}, has generated significant interest in alternative theories of gravity.
One such theory is massive gravity,  which involves assigning a mass to the graviton. This modification allows for a description of our universe's accelerating expansion without the need for a cosmological constant. In the framework of modern particle physics, the gravitational field can be understood as a theory of a spin-2 graviton\cite{Weinberg:1965pg,Boulware:1975cg}.
The concept of massive gravity dates back to 1939 when Fierz and Pauli constructed a linear theory of massive gravity \cite{Fierz:1939}. However, this theory suffers from a problem known as the Boulware-Deser ghost at the nonlinear level\cite{Boulware:1972if,Boulware:1972fr}. In recent years, a ghost-free massive theory, known as de Rham-Gabadadze-Tolley (dGRT) massive gravity\cite{Rham:2011tl}-\cite{Hinterbichler:2012}, was proposed, addressing the ghost issue.
In  dGRT  gravity theory, researchers have investigated a class of charged black holes and their thermodynamics in asymptotically AdS space-time\cite{Vegh:2013,Cai:2015tb}. The coefficients in the potential associated with the graviton mass have been found to play a similar role to the charge in thermodynamic phase space. Other black hole solutions have also been studied within the framework of massive gravity\cite{Xu:2015rfa}-\cite{Hendi:2018gb}.

In the low-energy limit of string theory, Einstein's gravity is recovered along with a non-minimal scalar dilaton field coupled to  axions and gauge fields~\cite{Green:1987sw}. In the context of coupling gravity to other gauge fields, the dilaton field is indispensable and plays a fundamental role in string theory. Its presence is necessary for the consistency of the theory. Moreover, the dilaton field has a significant impact on the causal structure of black holes. It induces modifications that result in curvature singularities occurring at finite radii \cite{Mignemi:1991wa}-\cite{Sheykhi:2007wg}.
There has been evidence that the dilaton potential, which is a generalization of the cosmological constant, modifies  the asymptotic behavior of solutions, making them neither flat nor (A)dS\cite{Gao:2004tu}-\cite{Dehyadegari:2017flm}.
Furthermore, exploring the coupling of the dilaton field to construct a unified theory of all interactions leads to scalar-tensor type generalizations of general relativity. These generalizations involve incorporating different types of curvature corrections into the standard Einstein-Hilbert Lagrangian coupled to the scalar field \cite{Berti:2015itd, Pani:2011xm, Yunes:2011we}.
One particular model that has received significant attention is the Einstein-dilaton-Gauss-Bonnet (EdGB) gravity~\cite{Kanti:1995vq,Torii:1996yi}, which incorporates a coupling function. Dilaton charges in this model are expressed as black hole masses, with a scalar dilaton field acting as a secondary hair. Black holes in various dimensions have been extensively studied within the framework of EdGB gravity \cite{Guo:2008hf, Ohta:2009tb, Ohta:2009pe, Kleihaus:2011tg, Maselli:2015tta}.

Recently, a scalar extension of dRGT massive gravity, known as massive Einstein-dilaton gravity\cite{Liu:2023sxw}, was explored. This extension involves considering the coupling of the dilaton field to the terms of massive graviton. Additionally, a specific class of static and spherically symmetric solutions for dilatonic black holes has been investigated.
Building upon these recent advancements, our objective is to go further beyond them and obtain the solutions for charged black holes in Maxwell-dilaton-massive gravity. In particular, we would like to highlight the following points of focus.
(i)The constructed Lagrangians consider the non-minimal coupling of the dilaton field solely to the terms of the graviton, as indicated in Eq.\eqref{eq4:action}. It is suggested that the coupling to the Maxwell field is minimal.
(ii)In contrast to the Ref.\cite{Liu:2023sxw}, we maintain an equal coupling strength of coupling for each term of the graviton unless otherwise specified for simplification.
(iii)Utilizing the derived Lagrangians, our aim is to obtain solutions for charged black holes within the framework of Maxwell-dilaton-massive gravity. Subsequently, we investigate the properties of these novel solutions and explore their unique characteristics.

The paper is organized as follows. In Sec.~\ref{2s}, we present the charged black hole solutions in four dimensional Maxwell-dilaton-massive gravity.  In Sec.~\ref{3s}, we further discuss the thermodynamic properties of these black holes in canonical and grand canonical ensembles.
Finally, a brief discussion is presented in Sec.~\ref{4s}.

\section{Solutions of charged black holes in Maxwell-dilaton-massive gravity}
\label{2s}
By considering the nonminimal  couplings of the dilaton field $\varphi$ to the graviton,
the action for
Maxwell-dilaton-massive gravity can be expressed   in four dimensional space-time as
\begin{equation}\label{eq4:action}
  I=\frac{1}{16 \pi} \int d^{4} x \sqrt{-g}
  \left(\mathcal {R}-F_{\mu\nu} F^{\mu\nu}  -2(\nabla \varphi)^2-V(\varphi)
  + e^{-2\beta \varphi}\mathcal {L}_m\right),
\end{equation}
where $V(\varphi)$ is a potential for the dilaton field $\varphi$,
and $F_{\mu\nu}$ is
the electromagnetic field tensor. The exponential factor of the last term  in Eq.(\ref{eq4:action}) denotes  the nonminimal coupling of the scalar dilaton field to the massive graviton with coupling constant $\beta$, and $\mathcal {L}_m$ is the so-called graviton action ~\cite{Rham:2011tl,Rham:2014mg}
\begin{equation}\label{eq:grav terms}
 \mathcal {L}_m= m_0^2 \sum_{i=1}^{4} \eta_i \mathcal {U}_i(g,h)
\end{equation}
where $m_0$ is the mass of the graviton, and $\eta_i$ are the dimensionless parameters used to distinguish different contributions of  $\mathcal{U}_i$.
Moreover, $h$ is a fixed rank-2 symmetric tensor,
and $\mathcal{U}_i$   satisfies the following recursion relation
\begin{eqnarray}\label{eq4:u1234}
  \mathcal{U}_1& =& [K]=K^{\mu}_{\mu}, \nonumber\\
  \mathcal{U}_2 &=& [K]^2 - [K^2], \nonumber\\
  \mathcal{U}_3 &=& [K]^3 - 3[K][K^2] + 2[K^3],\\
  \mathcal{U}_4 &=& [K]^4 - 6[K^2][K]^2 + 8[K^3][K] + 3[K^2]^2 - 6[K^4].\nonumber
  \end{eqnarray}
Here
$K^{\mu}_{\nu}=\sqrt{g^{\mu\tau}h_{\tau\nu}}$ .

Generally, the equations of motion can be obtained by computing the variation in action with respect to the field variables $g_{\mu\nu}$, $\varphi$ and $A_{\mu}$ as
\begin{eqnarray}
  G_{\mu\nu}&=&\mathcal{R}_{\mu\nu}-\frac{1}{2}\mathcal {R} g_{\mu\nu}\nonumber\\
                 &=& 2\partial_{\mu}\varphi\partial_{\nu}\varphi
                  -\frac{1}{2}(V + 2\partial^{\tau}\varphi\partial_{\tau}\varphi)g_{\mu\nu}+2(F_{\mu\tau}F_{\nu}^{\tau}-\frac{1}{4}F^2 g_{\mu\nu})
                 +m_0^2 e^{-2\beta \varphi} \chi_{\mu\nu}, \label{eq4:eist} \\
  \nabla^2\varphi  &=& \frac{1}{4}\big[\frac{\partial V}{\partial \varphi}
                  +2\beta e^{-2\beta \varphi}\mathcal {L}_m
                  \big],\label{eq4:dilaton}\\
   \nabla_\mu F^{\mu\nu}&=&0,\label{eq4:maxswell}
\end{eqnarray}
where
\begin{eqnarray}
\chi_{\mu\nu}&=& \frac{{\eta}_1}{2}(\mathcal{U}_1 g_{\mu\nu}-K_{\mu\nu})
  +\frac{{\eta}_2}{2}(\mathcal{U}_2 g_{\mu\nu}-2\mathcal{U}_1 K_{\mu\nu}
  +2K^2_{\mu\nu})\nonumber\\
  &+& \frac{{\eta}_3}{2}(\mathcal{U}_3 g_{\mu\nu}-3\mathcal{U}_2 K_{\mu\nu}
  +6\mathcal{U}_1 K^2_{\mu\nu}-6K^3_{\mu\nu})\nonumber \\
  &+& \frac{{\eta}_4}{2}(\mathcal{U}_4 g_{\mu\nu}-4\mathcal{U}_3 K_{\mu\nu}
  +12\mathcal{U}_2 K^2_{\mu\nu}-24\mathcal{U}_1 K^3_{\mu\nu}
  +24K^4_{\mu\nu})
  \label{eq4:chi}
\end{eqnarray}

The metric ansatz for a static spherically symmetric black hole solution takes the form
\begin{eqnarray}\label{eq4:metric}
  ds^2 = -f(r) dt^2 +f^{-1}(r) dr^2+ r^2R^2(r) d\Omega^2,
\end{eqnarray}
where $f(r)$ and $R(r)$ are functions of $r$, and $d\Omega^2=d\theta^2+\sin^2\theta d\phi^2$\ denotes the line element for two dimensional spherical  subspace.

Having supposed that the electromagnetic field has a component where only $A=A_t(r)dt $ for the static spherical symmetric  metric,
one can solve the Maxwell equation (\ref{eq4:maxswell}) 
\begin{equation}\label{eq4:em_tensor}
  F_{rt}=\frac{Q}{r^2 R^2}
\end{equation}
where $Q$ is the electric charge, making
use of Gauss's law  $Q=\frac{1}{4 \pi}\int {^*}F d\Omega$.

Distinguished from the dynamical physical metric $g_{\mu\nu}$, the reference metric $h_{\mu\nu}$ is usually fixed and assumed to be non-dynamical in massive theory.
Therefore, we  choose the reference metric to be\cite{Vegh:2013,Cai:2015tb,Xu:2015rfa}
\begin{eqnarray}\label{eq4:ref metr}
h_{\mu\nu}=diag(0, 0, c_0^2, c_0^2\sin^2\theta),
\end{eqnarray}
where $c_0$ is a positive parameter.
From the ansatz (\ref{eq4:ref metr}), the interaction potential in Eq.(\ref{eq4:u1234}) changes into
\begin{eqnarray}\label{u1:u2}
\mathcal{U}_1=\frac{2}{R r},\ \ \
\mathcal{U}_2=\frac{2}{R^2 r^{2}},\ \ \
\mathcal{U}_3=\mathcal{U}_4=0.
\end{eqnarray}
$\chi^{\mu}_{~\nu}$ in Eq.(\ref{eq4:chi}) can be written as
\begin{eqnarray}\label{eq4:chi:1}
\chi^1_{~1} = \chi^2_{~2} = \frac{e^{-2\beta \phi}( \eta_1 r R    + \eta_2 c_0^2 )}{(r R)^2 },\ \ \
\chi^3_{~3} = \chi^4_{~4} = \frac{\eta_1 e^{-2\beta \phi }}{2 r R }
\end{eqnarray}
and the components of equation of motion (\ref{eq4:eist}) become
\begin{eqnarray}
G^1_{~1}&=& \frac{1}{(r R)^2}[r R (r R)'f'+2r R (r R)''f+(r R)'^2 f-1] =-\frac{V(\varphi)}{2}-f \varphi'^2-\frac{Q^2}{r^4 R^4}+m_0^2 \chi^1_{~1}
\label{eq4:inseq:11},\\
  G^2_{~2} &=& \frac{1}{(r R)^2}[r R (r R)'f'+(r R)'^2 f -1 ]=-\frac{V(\varphi)}{2}+f \varphi'^2-\frac{Q^2}{r^4 R^4}+m_0^2 \chi^2_{~2}
  \label{eq4:inseq:22},\\
  G^3_{~3} &=& G^4_4=\frac{1}{ 2 r R}[(r R) f'' + 2 (r R)' f'+ 2 (r R)''f]=-\frac{V(\varphi)}{2}-f \varphi'^2+\frac{Q^2}{r^4 R^4}+m_0^2 \chi^3_{~3},
  \label{eq4:inseq:33}
\label{eq4:dilaton:1}
\end{eqnarray}
where the prime $'$ denotes differentiation with respect to $r$.

Calculating $G^1_{~1}-G^2_{~2}$, we obtain
 \begin{equation}\label{eq4:inseq:12}
   \frac{(r R)''}{r R}=-\varphi'^2,
 \end{equation}
which leads to
\begin{equation}\label{eq4:lnR}
   \frac{d^2}{dr^2}\ln R +\frac{2}{r}\frac{d}{dr}\ln R +\left(\frac{d}{dr}\ln R\right)^2=-\varphi'^2.
 \end{equation}
To obtain the dilaton field $\varphi$,  it is posited that $R(r)$ may be an exponential function of $\varphi(r)$
\begin{eqnarray}\label{eq4:ansatz}
  R(r)=e^{\alpha \varphi},
\end{eqnarray}
and Eq.(\ref{eq4:lnR}) reduces to
\begin{eqnarray}\label{eq4:dilaton:3}
\alpha \phi ''(r)+\left(\alpha ^2+1\right) \phi '(r)^2+\frac{2 \alpha  \phi '(r)}{r}=0,
\end{eqnarray}
where $\alpha$ is a positive constant. Then, the dilaton field can be obtained as
\begin{equation}\label{eq4:varphi}
  \varphi (r) = \frac{\alpha}{1+\alpha^2} \ln\frac{\delta}{r}.
\end{equation}
Here $\delta$ is a constant parameter.

Regarding these solutions together with the electromagnetic
field tensor Eq.(\ref{eq4:em_tensor}),
we can obtain
\begin{eqnarray}\label{eq4:at}
 A_t(r)=
 \left\{  \begin{array}{ll}
  \frac{(\alpha ^2+1) Q} {\alpha ^2-1}\delta ^{-\frac{2 \alpha ^2}{\alpha ^2+1}} r^{-\frac{1-\alpha ^2}{1+\alpha ^2}}, &
    \alpha\neq1\\     \frac{Q \log (r)}{\delta }, &  \alpha =1  \end{array}    \right.
\end{eqnarray}

Note that  the potential function
$A_{t}(r)$ should be vanishing at infinity  to be physically reasonable~\cite{KordZangeneh:2015fdy},  therefore the condition $\alpha<1$ must be fulfilled and the solution of  $A_{t}(r)$ is only the first branch in Eq.\eqref{eq4:at}.

According to the metric ansatz (\ref{eq4:metric}) and expressions for 
dilaton field (\ref{eq4:ansatz})(\ref{eq4:varphi}), the Klein-Gordon equation (\ref{eq4:dilaton}) becomes
\begin{eqnarray}
\frac{f'}{\alpha ^2+1}+\frac{\left(1-\alpha ^2\right) f}{(\alpha ^2 +1)^2 r}
  +\frac{ r}{4 \alpha }\frac{\partial V}{\partial \varphi}
  +\frac{  m_0^2\beta c_0 e^{-2\beta \varphi}}{\alpha \delta}
  \left(\eta_1 \delta e^{-\alpha \varphi}+\eta_2 c_0 e^{\frac{1-\alpha^2}{\alpha} \varphi} \right)=0.
 \label{eq4:dilaton:2}
\end{eqnarray}
Moreover, we further consider the trace of the gravitational equation (\ref{eq4:eist})
\begin{eqnarray}\label{eq4:contract}
f''&+&\frac{4( r R)'f'}{rR}+\frac{ 2 f\left((r R')^2+2 r R \left(r R''+3 R'\right)+R^2+(r R)^2\varphi '^2 \right)}{(r R)^2}\nonumber\\
&+&2 V(\varphi)-\frac{2}{(r R)^2}-m_0^2\sum^4_{i=1}\chi_i^i=0.
\end{eqnarray}
Based on $G^3_{~3}$ component of gravitational equation Eq.(\ref{eq4:inseq:33}), Eq.\eqref{eq4:contract} can be written as 
\begin{eqnarray}  \label{eq4:inseq:33:2}
 -\frac{f'}{\alpha ^2+1}
   +\frac{\left(\alpha   ^2-1\right) f}{\left(\alpha ^2+1\right)^2 r}
   -\frac{ r V(\varphi)}{2}
   +\frac{e^{\frac{1-\alpha^2}{\alpha}\varphi } }{\delta}
   -\frac{ Q^2}{\delta^3} e^{\frac{3- \alpha ^2}{\alpha}\varphi}\nonumber\\
 + \frac{m_0^2 c_0 e^{-2\beta \varphi}}{\delta}\left(\eta_1 \delta e^{-\alpha\varphi }
+\eta_2 c_0e^{\frac{1-\alpha^2}{\alpha} \varphi}\right)
=0
\end{eqnarray}
together with the expressions for dilaton field (\ref{eq4:varphi}).

Considering Eqs.(\ref{eq4:dilaton:2})(\ref{eq4:inseq:33:2}), we can obtain a first order differential equation for dilaton field potential
\begin{eqnarray}
&&\frac{ \partial V(\varphi )}{\partial \varphi }-2 \alpha  V(\varphi  )
+\frac{4 \alpha  e^{\frac{2 \varphi  }{\alpha }}}{\delta ^2}
-\frac{4 \alpha  Q^2 e^{\frac{4 \varphi  }{\alpha }}}{\delta ^4}\nonumber\\
&&+\frac{4 \eta_1 c_0 m_0^2 (\alpha +\beta ) e^{-2 \varphi   \left(\beta -\frac{1}{2 \alpha }\right)}}{\delta }
+\frac{4 \eta_2 c_0^2 m_0^2 (\alpha +\beta ) e^{-2 \varphi  \left(\beta -\frac{1}{\alpha }\right)}}{\delta ^2}
=0,
\label{eq4:potential :dilation}
\end{eqnarray}
which leads to
\begin{eqnarray}\label{eq4:di potential}
  V(\varphi)=&&
  \frac{2 \alpha ^2 e^{\frac{2 \varphi }{\alpha }}}{\left(\alpha ^2-1\right) \delta ^2}
  +2 \Lambda  e^{2 \alpha  \varphi }
  -\frac{2 \alpha ^2 Q^2 e^{\frac{4 \varphi }{\alpha }}}{\left(\alpha ^2-2\right) \delta ^4}\nonumber\\
  &&+\frac{4 \alpha  c_0 \eta_1 m_0^2 (\alpha +\beta ) e^{\varphi  \left(\frac{1}{\alpha }-2 \beta \right)}}{\delta  \left(2 \alpha ^2+2 \alpha  \beta -1\right)}\nonumber\\
  &&+\frac{2 \alpha  c_0^2 \eta_2 m_0^2 (\alpha +\beta ) e^{2 \varphi  \left(\frac{1}{\alpha }-\beta \right)}}{\delta ^2 \left(\alpha ^2+\alpha  \beta -1\right)}.
\end{eqnarray}
Here the last two terms are associated with the coupling between the dilaton field and the graviton, and $\Lambda$ is a free parameter that is considered to
play the role of the cosmological constant, because in the absence
of the dilaton field ($\alpha=0$), we have $V=2\Lambda$,
i.e. $\Lambda=-\frac{d(d-1)}{2l^2}$, in which $l$ is the radius of the $(d+1)-$ dimensional AdS spacetime.
Then, the metric function $f(r)$ from Eq.~(\ref{eq4:dilaton:2}) can be also obtained as
\begin{eqnarray}\label{eq4:metr fina}
  f(r)=
  &-m r^{\frac{\alpha ^2-1}{\alpha ^2+1}}
  -\frac{\left(\alpha ^2+1\right) \left(\frac{\delta }{r}\right)^{-\frac{2 \alpha ^2}{\alpha ^2+1}}}{\alpha ^2-1}
  +\frac{2 \left(\alpha ^2+1\right)^2 Q^2 \delta ^{-\frac{4 \alpha ^2}{\alpha ^2+1}} r^{\frac{2 \left(\alpha ^2-1\right)}{\alpha ^2+1}}}{\alpha ^4-3 \alpha ^2+2}
  +\frac{\left(\alpha ^2+1\right)^2 \Lambda  \delta ^{\frac{2 \alpha ^2}{\alpha ^2+1}} r^{\frac{2}{\alpha ^2+1}}}{\alpha ^2-3}\nonumber\\
  &-\frac{\left(\alpha ^2+1\right)^2 c_0 \eta_1 m_0^2 \delta ^{-\frac{\alpha  (\alpha +2 \beta )}{\alpha ^2+1}} r^{\frac{2 \alpha ^2+2 \alpha  \beta +1}{\alpha ^2+1}}}{\left(\alpha ^2+2 \alpha  \beta +2\right) \left(2 \alpha ^2+2 \alpha  \beta -1\right)}
  -\frac{\left(\alpha ^2+1\right)^2 c_0^2 \eta_2 m_0^2 \left(\frac{\delta }{r}\right)^{-\frac{2 \alpha  (\alpha +\beta )}{\alpha ^2+1}}}{\left(\alpha ^2+\alpha  \beta -1\right) \left(\alpha ^2+2 \alpha  \beta +1\right)}
\end{eqnarray}
where $m$ is an integration constant related to the mass of the black hole as it will be shown below.
In the absence of the dilaton field $(\alpha=0)$, the metric function $f(r)$ becomes
\begin{eqnarray}
f(r)=1-\frac{m}{r}+\frac{ Q^2
}{ r^{2}}+\frac{1}{2} \eta_1m_0^2 r+\eta_2 m_0^2-\frac{1}{3} \Lambda  r^2,
\end{eqnarray}
which was presented in Ref.\cite{Cai:2015tb}\
(The value of $Q$ as $\alpha\rightarrow0$ is a half of  the one in Ref.\cite{Cai:2015tb}) due to  the different factors  of the electromagnetic action $F^2$). 
Here we have set $c_0=1$.
Obviously, the metric function $f(r)$ does not describe an AdS space unless $m_0=0$.

\begin{figure}[H]
  \subfigure[]{\label{fig:1:1} 
  \includegraphics[width=5cm]{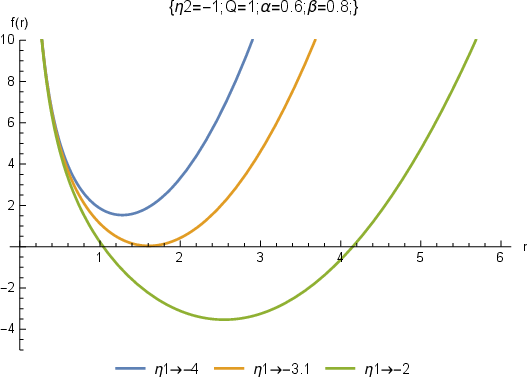}}%
  \subfigure[]{\label{fig:1:2} 
  \includegraphics[width=5cm]{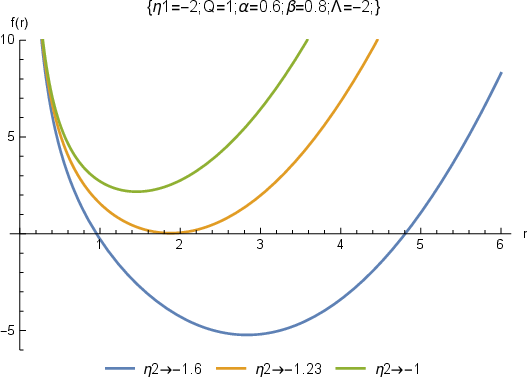}}%
  \subfigure[]{\label{fig:1:3} 
  \includegraphics[width=5cm]{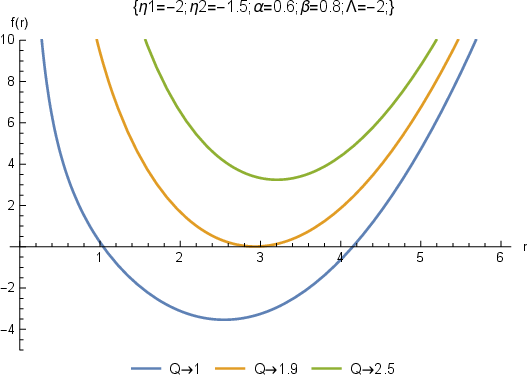}}\\
  \subfigure[]{\label{fig:1:4} 
  \includegraphics[width=5cm]{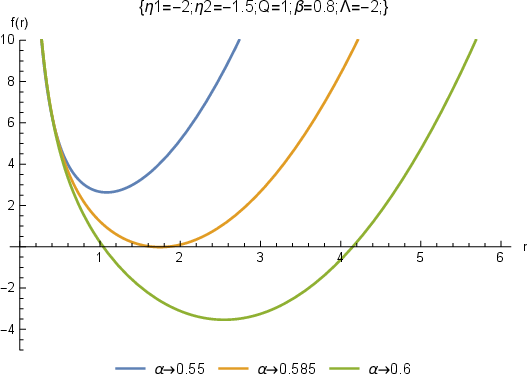}}%
  \subfigure[]{\label{fig:1:5} 
  \includegraphics[width=5cm]{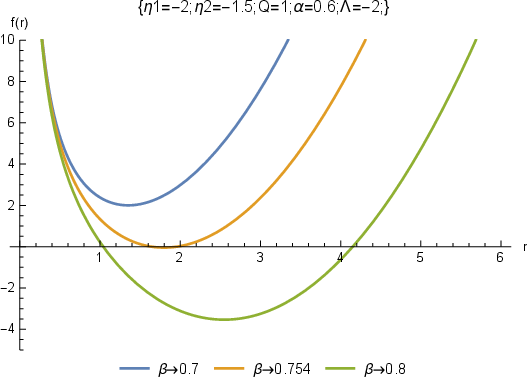}}%
  \caption{Diagrams of metric function $f(r)$ versus $r$ for $m_0=c_0=\delta=m=1$,and $\Lambda=-2$}
  \label{fig:1}
\end{figure}

At infinity, the dominant term of the solution $f(r)$ approaches
\begin{eqnarray} \label{eq:asymp}
\lim_{r\to \infty} f(r)=
  \frac{\left(\alpha ^2+1\right)^2 \Lambda  \delta ^{\frac{2 \alpha ^2}{\alpha ^2+1}} r^{\frac{2}{\alpha ^2+1}}}{\alpha ^2-3}
  -\frac{\left(\alpha ^2+1\right)^2 c_0 \eta_1 m_0^2 \delta ^{-\frac{\alpha  (\alpha +2 \beta )}{\alpha ^2+1}} r^{\frac{2 \alpha ^2+2 \alpha  \beta +1}{\alpha ^2+1}}}{\left(\alpha ^2+2 \alpha  \beta +2\right) \left(2 \alpha ^2+2 \alpha  \beta -1\right)}
\end{eqnarray}
For example,
taking $\alpha =\beta=0.6$, we have
\begin{equation}
    \lim_{r\to \infty} f(r)=-1.36482 m_0^2 c_0 \eta_1 r^{1.79412}- 0.700606 \Lambda r^{1.47059},
\end{equation}
where we set $\delta=1$ for simplicity. Clearly, the metric function is  neither asymptotically flat nor asymptotically (A)dS.
To comprehensively understand the behavior of the metric function, we graphically present the  dependence of the function $f(r)$.
Plots of metric function versus $r$ are shown in Fig.\ref{fig:1} in terms of different parameters, which show that two horizon, extreme and naked singularity black holes can occur by choosing the proper parameters.

Now we consider the spacetime singularities, and calculate the Kretschmann scalar
\begin{eqnarray}
  \mathcal {R}^{\mu\nu\rho\sigma} \mathcal {R}_{\mu\nu\rho\sigma}
   &=&f \left(\frac{8 f' (r R)' (r R)''}{(r R)^2}-\frac{8 \left((r R)'\right)^2}{(r R)^4}\right)+\frac{4 f'^2 \left((r R)'\right)^2}{(r
   R)^2}\nonumber\\
   & &+\frac{4 f^2 \left(\left((r R)'\right)^4+2 (r R)^2 \left((r R)''\right)^2\right)}{(r R)^4}+\frac{4}{(r R)^4}+f''^2.
\end{eqnarray}

Evidently, the Kretschmann scalar is not singular at the horizons. These points are simple singularity of coordinate as for a black hole. From the leading terms of asymptotic behaviors of metric at the origin, we have
\begin{eqnarray}
 \lim_{r\rightarrow0} \mathcal {R}^{\mu\nu\rho\sigma}\mathcal {R}_{\mu\nu\rho\sigma}=\infty,\ \
 \lim_{r\rightarrow\infty} \mathcal {R}^{\mu\nu\rho\sigma}\mathcal {R}_{\mu\nu\rho\sigma}=0,
\end{eqnarray}
where the Kretschmann scalar $\mathcal {R}^{\mu\nu\rho\sigma}\mathcal {R}_{\mu\nu\rho\sigma}$ is divergence  at
origin $r=0$, finite for $r>0$, and vanishes at infinity,
which suggests the  origin $r=0$  is an essential and  physical singularity in the spacetime.

\section{Thermodynamics and critical behaviors of black holes}
\label{3s}

In this section, we discuss the thermodynamic properties of black holes from the previous section. The black hole entropy can be defined using well-established Wald's approach\cite{Wald:1993nt,Iyer:1994ys}:
\begin{equation}
  S=-2\pi \int_{\Sigma} d^{d-2}x \sqrt{\sigma}\frac{\partial \mathcal {L}}{\partial\mathcal {R}_{\mu\nu\kappa\lambda}}\varepsilon_{\mu\nu}\varepsilon_{\kappa\lambda}, 
\end{equation}
where $\Sigma$ denotes a Killing horizon with binormal $\varepsilon_{\mu\nu}$,
$\sigma$ is the determinant of the induced metric on the horizon surface, and $\mathcal {L}$ is the
gravitational Lagrangian.
Therefore, the entropy of these black holes is
\begin{eqnarray}  \label{eq4:entropy}
S=\pi  \delta ^{\frac{2 \alpha ^2}{\alpha ^2+1}} r_h^{\frac{2}{\alpha ^2+1}}.
\end{eqnarray}
This is identified with the  so-called  entropy-area law. 
Refs.~\cite{Cai:2015tb}-\cite{Zou:2016sab} show that graviton mass has only a small effect on the form of entropy, and is just  a correction for the horizon radius.

Moreover, the temperature of black holes in terms of the null Killing vector $(\frac{\partial}{\partial t})^a$ at the horizon reads as
\begin{eqnarray}\label{eq4:temp}
T &=& \frac{\kappa}{2 \pi}=\frac{1}{4 \pi}\frac{\partial f(r)}{\partial r}\big|_{r=r_h} \nonumber\\
  &=&
   \frac{\left(\alpha ^2+1\right) Q^2 \delta ^{-\frac{4 \alpha ^2}{\alpha ^2+1}} r_h^{\frac{\alpha ^2-3}{\alpha ^2+1}}}{2 \pi  \left(\alpha ^2-2\right)}
  -\frac{\left(\alpha ^2+1\right) \Lambda  \delta ^{\frac{2 \alpha ^2}{\alpha ^2+1}} r_h^{\frac{2}{\alpha ^2+1}-1}}{4 \pi }
  -\frac{\left(\alpha ^2+1\right) \left(\frac{\delta }{r_h}\right){}^{-\frac{2 \alpha ^2}{\alpha ^2+1}}}{4 \pi  \left(\alpha ^2-1\right) r_h}
  \nonumber\\
  & &-\frac{\left(\alpha ^2+1\right) c_0 \eta _1 m_0^2 \left(\frac{r_h}{\delta }\right){}^{\frac{\alpha  (\alpha +2 \beta )}{\alpha ^2+1}}}{4 \pi  \left(2 \alpha ^2+2 \alpha  \beta -1\right)}
  -\frac{\left(\alpha ^2+1\right) c_0^2 \eta _2 m_0^2 \left(\frac{\delta }{r_h}\right){}^{-\frac{2 \alpha  (\alpha +\beta )}{\alpha ^2+1}}}{4 \pi  \left(\alpha ^2+\alpha  \beta -1\right) r_h}.
\end{eqnarray}
From the definition of ADM mass~\cite{Abbott:1982},
we obtain 
\begin{eqnarray}\label{eq4:mass}
  M&=&\frac{\delta^{\frac{2\alpha^2}{1+\alpha^2}}m}{2(1+\alpha^2)}\nonumber\\
&=&\frac{\left(\alpha ^2+1\right) Q^2 \delta ^{-\frac{2 \alpha ^2}{\alpha ^2+1}} r_h^{\frac{\alpha ^2-1}{\alpha ^2+1}}}{\alpha ^4-3 \alpha ^2+2}
+\frac{\left(\alpha ^2+1\right) \Lambda  \delta ^{\frac{4 \alpha ^2}{\alpha ^2+1}} r_h^{\frac{3-\alpha ^2}{\alpha ^2+1}}}{2 \left(\alpha ^2-3\right)}
-\frac{\left(\frac{1}{r_h}\right){}^{-\frac{2 \alpha ^2}{\alpha ^2+1}} r_h^{\frac{1-\alpha ^2}{\alpha ^2+1}}}{2 \left(\alpha ^2-1\right)}\nonumber\\
& &-\frac{\left(\alpha ^2+1\right) c_0 \eta _1 m_0^2 \delta ^{\frac{\alpha  (\alpha -2 \beta )}{\alpha ^2+1}} r_h^{\frac{\alpha ^2+2 \alpha  \beta +2}{\alpha ^2+1}}}{2 \left(\alpha ^2+2 \alpha  \beta +2\right) \left(2 \alpha ^2+2 \alpha  \beta -1\right)}
-\frac{\left(\alpha ^2+1\right) c_0^2 \eta _2 m_0^2 \delta ^{\frac{2 \alpha ^2}{\alpha ^2+1}} r_h^{\frac{1-\alpha ^2}{\alpha ^2+1}} \left(\frac{\delta }{r_h}\right){}^{-\frac{2 \alpha  (\alpha +\beta )}{\alpha ^2+1}}}{2 \left(\alpha ^2+\alpha  \beta -1\right) \left(\alpha ^2+2 \alpha  \beta +1\right)}
\end{eqnarray}

The electric potential $\Phi$ of black holes, measured by an observer located at infinity with respect
to the horizon, can be calculated using the following standard relation\cite{KordZangeneh:2015hfy}
\begin{eqnarray} \label{eq4:elecpot def}
\Phi=A_{\mu} Z^{\mu}|_{r\rightarrow \infty}-A_{\mu} Z^{\mu}|_{r=r_h},
\end{eqnarray}
where $Z = C \partial_t$ is the null generator of the horizon.Therefore, using Eq.\eqref{eq4:at} the electric potential may be obtained as
\begin{eqnarray} \label{eq4:elecpot 1}
\Phi=
  -\frac{C (\alpha ^2+1) Q \delta ^{-\frac{2 \alpha ^2}{\alpha ^2+1}} r^{-\frac{1-\alpha ^2}{1+\alpha ^2}}}{\alpha ^2-1}.
\end{eqnarray}

Regarding the parameters $S$ and $Q$ as a complete set of extensive parameters,  
 the Smarr-type mass formula for the new black holes is expressed as
\begin{eqnarray}  \label{eq4:smarr mass}
M(S,Q)&=& \frac{\left(\alpha ^2+1\right) Q^2 \delta ^{-\alpha ^2} (\frac{S}{\pi})^{\frac{1}{2} \left(\alpha ^2-1\right)}}{\alpha ^4-3 \alpha ^2+2}
+\frac{\delta ^{-\alpha ^2} (\frac{S}{\pi})^{\frac{1}{2} \left(\alpha ^2+1\right)}}{2-2 \alpha ^2}\nonumber\\
& &
-\frac{\left(\alpha ^2+1\right) c_0 \eta_1 m_0^2 \delta ^{-\alpha  (\alpha +2 \beta )} (\frac{S}{\pi})^{\frac{\alpha ^2}{2}+\alpha  \beta +1}}{2 \left(\alpha ^2+2 \alpha  \beta +2\right) \left(2 \alpha ^2+2 \alpha  \beta -1\right)}
\nonumber\\
& &
-\frac{\left(\alpha ^2+1\right) c_0^2 \eta_2 m_0^2 \delta ^{-\alpha  (\alpha +2 \beta )} (\frac{S}{\pi})^{\frac{1}{2} \left(\alpha ^2+2 \alpha  \beta +1\right)}}{2 \left(\alpha ^2+2 \alpha  \beta +1\right) (\alpha  (\alpha +\beta )-1)}
+\frac{\left(\alpha ^2+1\right) \Lambda  \delta ^{\alpha ^2} (\frac{S}{\pi})^{\frac{3}{2}-\frac{\alpha ^2}{2}}}{2 \left(\alpha ^2-3\right)}
\end{eqnarray}
One may find that the intensive parameters $T$  and $\Phi$, conjugate to the
black hole entropy and charge, satisfy the following relations
\begin{eqnarray}  \label{eq4:conjugate}
T=\left(\frac{\partial M(S,Q)}{\partial S}\right)_Q,\ \ \
\Phi=\left(\frac{\partial M(S,Q)}{\partial Q}\right)_S,
\end{eqnarray}
provided that $C$ should be chosen as
\begin{eqnarray}  \label{eq4:c}
C=\frac{2}{2-\alpha ^2}.
\end{eqnarray}
Note that
$ C= 1$ and $\Phi=\frac{ Q }{r}$ in the case of $\alpha=0$. It is shown that the intensive quantities calculated by Eq.\eqref{eq4:conjugate}, regarded as the temperature and electric potential,     coincide with Eqs.\eqref{eq4:temp} and Eqs.\eqref{eq4:elecpot 1}
. Thus,  these thermodynamics quantities satisfy the first law of the thermodynamics  expressed as
\begin{eqnarray}\label{eq: first law}
  d M = T dS + \Phi dQ
\end{eqnarray}

Then we would like to investigate the critical behaviors of the charged black holes both in grand canonical ensemble and in canonical ensemble,
and we will set $m_0=c_0=\delta=1$ and $\Lambda=-2$ for simplification in the following discussions.

\subsection{ Critical behaviors of black holes in grand canonical ensemble}

In grand canonical ensemble with a fixed chemical potential $\Phi$ associated with the charge $Q$,
The Gibbs free energy is written as
\begin{eqnarray}\label{eq4:gib}
  G&=&M-TS-\Phi Q\nonumber\\
  &=&\frac{r_h}{4}-\frac{1}{8} \left(\alpha ^4-3 \alpha ^2+2\right) \Phi ^2 r_h^{\frac{1-\alpha ^2}{\alpha ^2+1}}
  +\frac{\left(\alpha ^2-1\right) \left(\alpha ^2+1\right) \Lambda  r_h^{\frac{3-\alpha ^2}{\alpha ^2+1}}}{4 \left(\alpha ^2-3\right)}\\
  & &+\frac{\left(\alpha ^2+1\right) \eta _2 \left(\alpha ^2+2 \alpha  \beta -1\right) r_h^{\frac{\alpha ^2+2 \alpha  \beta +1}{\alpha ^2+1}}}{4 \left(\alpha ^2+\alpha  \beta -1\right) \left(\alpha ^2+2 \alpha  \beta +1\right)}
  +\frac{\alpha  \left(\alpha ^2+1\right) \eta _1 (\alpha +2 \beta ) r_h^{\frac{\alpha ^2+2 \alpha  \beta +2}{\alpha ^2+1}}}{4 \left(\alpha ^2+2 \alpha  \beta +2\right) \left(2 \alpha ^2+2 \alpha  \beta -1\right)}
  \nonumber
\end{eqnarray}

The temperature of black hole is
\begin{eqnarray}\label{eq:temp:gra}
 T&=&
 \frac{\left(\alpha ^2-2\right) \left(\alpha ^2-1\right)^2 \Phi ^2}{8 \pi  \left(\alpha ^2+1\right) r_h}
 -\frac{\left(\alpha ^2+1\right) \Lambda  r_h^{\frac{2}{\alpha ^2+1}-1}}{4 \pi }
 -\frac{\left(\alpha ^2+1\right) r_h^{\frac{\alpha ^2-1}{\alpha ^2+1}}}{4 \pi  \left(\alpha ^2-1\right)}\\
 & &-\frac{\left(\alpha ^2+1\right) \eta _1 r_h^{\frac{\alpha  (\alpha +2 \beta )}{\alpha ^2+1}}}{4 \pi  \left(2 \alpha ^2+2 \alpha  \beta -1\right)}
 -\frac{\left(\alpha ^2+1\right) \eta _2 r_h^{\frac{2 \alpha  (\alpha +\beta )}{\alpha ^2+1}-1}}{4 \pi  \left(\alpha ^2+\alpha  \beta -1\right)}
 \nonumber
\end{eqnarray}
and the corresponding heat capacity is
\begin{eqnarray}
 C_\Phi=\frac{-16 \pi ^2 T r^{\frac{2}{\alpha ^2+1}+1}(\alpha ^4-1)^{-1} }{\frac{\left(\alpha ^4-3 \alpha ^2+2\right) \Phi ^2}{\alpha ^2+1}+\frac{2 \eta _2 \left(\alpha ^2+2 \alpha  \beta -1\right) r_h^{\frac{2 \alpha  (\alpha +\beta )}{\alpha ^2+1}}}{\left(\alpha ^2-1\right) \left(\alpha ^2+\alpha  \beta -1\right)}+\frac{2 \alpha  \eta _1 (\alpha +2 \beta ) r_h^{\frac{\alpha  (\alpha +2 \beta )}{\alpha ^2+1}+1}}{\left(\alpha ^2-1\right) \left(2 \alpha ^2+2 \alpha  \beta -1\right)}-2 \Lambda  r_h^{\frac{2}{\alpha ^2+1}}+\frac{2 r_h^{\frac{2 \alpha ^2}{\alpha ^2+1}}}{\alpha ^2-1}}
\end{eqnarray}

According to Eq.(\ref{eq4:gib}),it is worth noting that, the term of $\eta_1$ will vanish for $\alpha\rightarrow 0$,
which means that the term of $\eta_2$ is much more important in the case without the dilaton field.
But in the case with the dilaton field, i.e.$\alpha\neq 0$,
considering the asymptotic behaviors of the Gibbs free energy,
the term of $\eta_1$ play more important role than the one of $\eta_2$ as $r_h\rightarrow \infty$.

 Using the conditions of inflection point in Van der Waals system
\begin{eqnarray}
  \frac{\partial T}{\partial r_h}\left|_{\Phi=\Phi_c,r_h=r_c}\right.
  =\frac{\partial^2 T}{\partial r_h^2}\left|_{\Phi=\Phi_c,r_h=r_c}\right.
  =0,
\end{eqnarray}
we can receive the critical $\Phi_c$ and the equation of
critical radius of black hole as
\begin{align}
 \Phi^2_{c}= &\frac{2 \left(\alpha ^2+1\right)}{\left(\alpha ^2-2\right) \left(\alpha ^2-1\right)^2}\left(
 \left(\alpha ^2-1\right) \Lambda  r_c^{\frac{2}{\alpha ^2+1}} 
 -\frac{\eta_2 \left(\alpha ^2+2 \alpha  \beta -1\right) }{\alpha ^2+\alpha  \beta -1}\cdot  r_c^{\frac{2 \alpha  (\alpha +\beta )}{\alpha ^2+1}}
 \right.\nonumber\\
 &\left.  -r_c^{\frac{2 \alpha ^2}{\alpha ^2+1}}
 -\frac{\alpha  \eta_1 (\alpha +2 \beta ) }{2 \alpha ^2+2 \alpha  \beta -1}\cdot r_c^{\frac{\alpha  (\alpha +2 \beta )}{\alpha ^2+1}+1}
 \right)
 \label{eq:phic}\\
  0= & \alpha ^2 +\left(1-\alpha ^2\right) \Lambda  r_c^{\frac{2-2 \alpha ^2}{\alpha ^2+1}}+\frac{\alpha  \eta _2 (\alpha +\beta ) \left(\alpha ^2+2 \alpha  \beta -1\right) }{\alpha ^2+\alpha  \beta -1}\cdot r_c^{\frac{2 \alpha  \beta }{\alpha ^2+1}} \nonumber\\
 &+\frac{\alpha  \eta _1 (\alpha +2 \beta ) \left(2 \alpha ^2+2 \alpha  \beta +1\right) }{4 \alpha ^2+4 \alpha  \beta -2}\cdot r_c^{\frac{2 \alpha  \beta +1}{\alpha ^2+1}} \label{eq:g cri radi}
\end{align}
When $\alpha=0$, there are no critical radii satisfying the above equation \eqref{eq:g cri radi}. It means that there are no VdW-like phase transitions without  dilaton field. For $\alpha\neq 0$,
there could be some critical radii of the Eq.\eqref{eq:g cri radi}, which implies that it exists more critical phenomena of this system in grand
canonical ensemble.
But it is necessary to note that there are several conditions should be considered when we study these critical behaviors.
Firstly, the roots of Eq.\eqref{eq:g cri radi}, which are radii of black holes, must be positive. Secondly, From Eq.\eqref{eq:phic}, the right-hand of the equation $\Phi^2_{c}$
must be positive. Finally,   the temperatures of the black holes $T_c$  with respect to the critical radii must also be positive.
Basing on the three conditions,  so we just numerically analyze the critical behaviors of the system, and find that there are at most two critical radii for Eq.(\ref{eq:g cri radi}).
Taking the case $ \alpha = 0.4$, and $\beta = 1.6$ for example, there could be at most two critical radii  satisfying the positive temperature for $\eta_1 >0 $  but only one for $\eta_1 <0$.

\begin{figure}[H]
  \subfigure[$G-T$ diagram with  $\Phi^2$ ]{\label{fig:grand1:1} 
  \includegraphics[width=8cm]{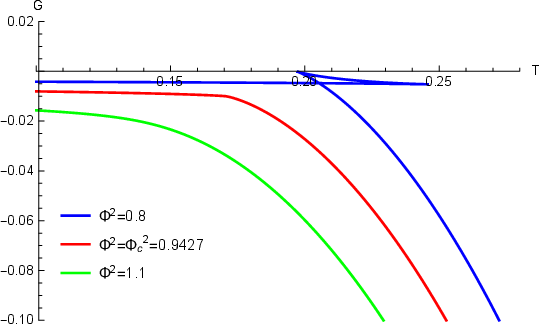}}%
  \hfill%
  \subfigure[$T-\Phi^2$ diagram ]{\label{fig:grand1:2} 
  \includegraphics[width=8cm]{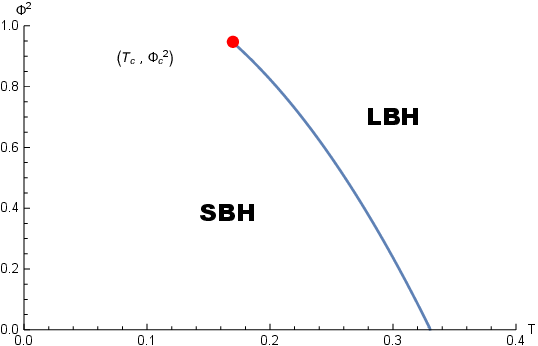}}
  \caption{Critical behaviors for $\eta_1=-1$, $\eta_2=-0.2$, $\alpha=0.4$, $\beta =1.6$}.
  \label{fig:grand1}
\end{figure}

In the case one critical radius,
there is a first-order VdW-like phase transition.
In the $G-T$ diagram in Fig.\ref{fig:grand1:1},
the red line corresponds to critical point at $\Phi^2_{c}\approx0.94270$,
and it appears  a ``swallowtail" in when $\Phi^2<\Phi^2_{c}$.
The phase diagram is displayed in Fig.\ref{fig:grand1:2},
and the coexistence line terminates at the critical point ($Tc\approx0.17062$, $\Phi^2_{c}\approx0.94270$) as increasing the temperature .

\begin{figure}[H]
  \subfigure[  ]{\label{fig:grandgt:1} 
  \includegraphics[width=8cm]{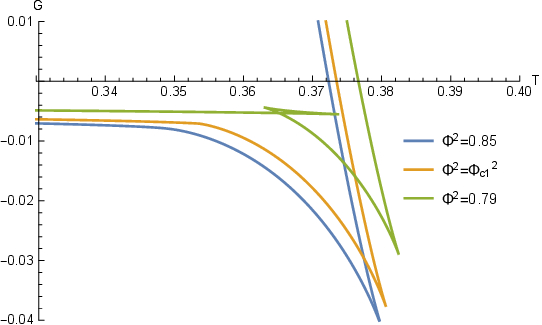}}%
  \subfigure[ ]{\label{fig:grandgt:2} 
  \includegraphics[width=8cm]{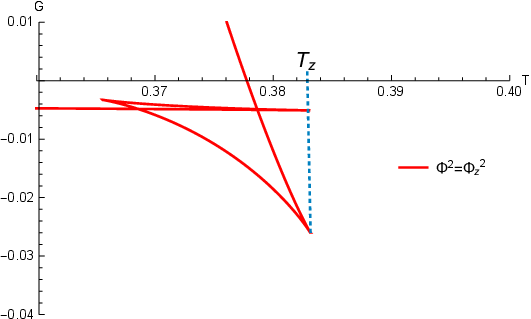}}%
\\
  \subfigure[ ]{\label{fig:grandgt:3} 
  \includegraphics[width=8cm]{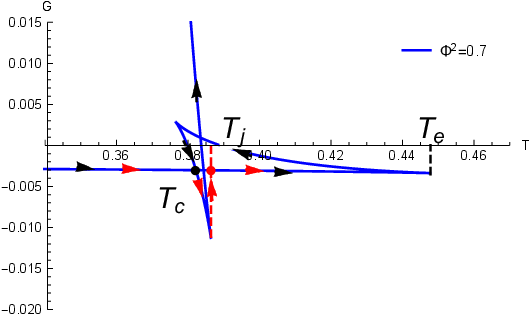}}
  \subfigure[]{\label{fig:grandgt:4} 
  \includegraphics[width=8cm]{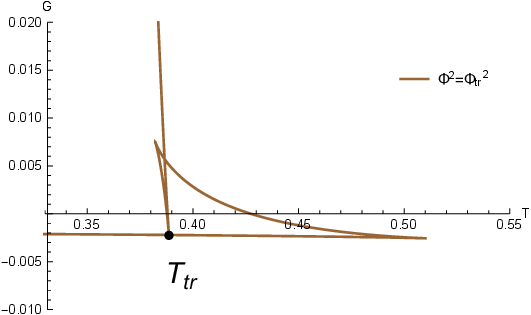}}%
  \\
  \subfigure[ ]{\label{fig:grandgt:5} 
  \includegraphics[width=8cm]{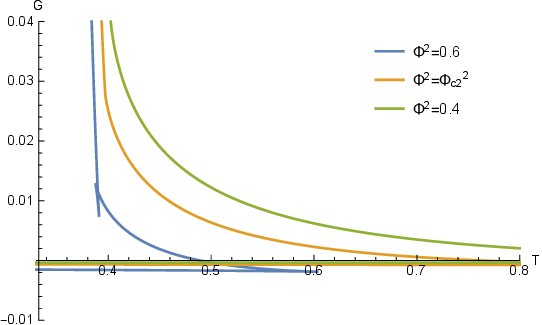}}%
  \subfigure[]{\label{fig:grandgt:6} 
  \includegraphics[width=8cm]{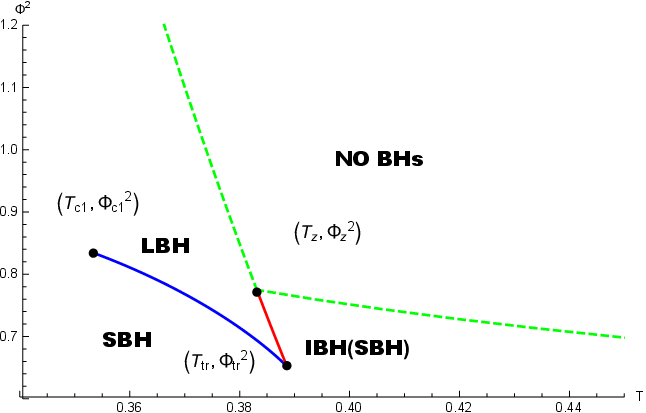}}
  \caption{Diagrams with two critical radii for $\eta_1= 1$, $\eta_2=0.6$, $\alpha=0.4$, and $\beta=1.6$ }
    \label{fig:grandgt}
\end{figure}

In the case of two critical radii, Figs.\ref{fig:grandgt:1}-\ref{fig:grandgt:5} depict the critical behaviors of the Gibbs free energy $G$. Through a detailed investigation of the processes, we observe a triple critical point $ ( T_{tr}, {\Phi}_{tr}^2) $ and two types of phase transitions: zero and first-order.
It is important to emphasize that these phase transitions are novel and intriguing as they exhibit a reversal of the reentrant phase transitions\cite{Zou:2016sab}. This reversal is illustrated as follows.
When $\Phi^2_{z}<\Phi^2<\Phi^2_{c1}$, as shown in Fig.\ref{fig:grandgt:1}, we observe a ``swallowtail" pattern, indicating the occurrence of a  VdW phase transition owing to the global minimum value of the Gibbs free energy. The critical point ($\Phi^2_{z}$, $T_{z}$)is depicted in Fig.\ref{fig:grandgt:2}.
In the range $\Phi^2_{tr}<\Phi^2<\Phi^2_{z}$  (see Fig.\ref{fig:grandgt:3}), the black hole radius increases along the line from left to right. As the temperature increases from left to right, the value of the free energy, $G$, decreases along the lower line, considering the minimum of the global Gibbs free energy. The red arrows highlight the physical process of the system with respect to temperature.
A swallowtail pattern with a black cross point ($T=T_c$) indicates a first-order phase transition from a small black hole (SBH) to a large black hole (LBH). As the temperature continues to increase, the value of the Gibbs free energy, exhibits a discontinuous ``jump-up" to the upper line at $T=T_j$, signifying a zero-order phase transition from an LBH to an SBH (also referred to as an intermediate black hole (IBH) because it is larger than the former SBH).
The process concludes at the point $T=T_e$, beyond which there are no black holes for $T>T_e$. Thus, the system undergoes a novel ``SBH-LBH-SBH" phase transition process, which is the reverse of the ``LBH-SBH-LBH" process observed in reentrant phase transitions. Owing to these characteristics, we refer to this process as a ``reverse" reentrant phase transition.

To provide a clearer explanation, we can refer to the phase diagram shown in Fig. \ref{fig:grandgt:6}. In this diagram, the blue line represents the coexistence line between small black holes (SBH) and large black holes (LBH), while the red line represents the coexistence line between LBH and intermediate black holes (IBH).
The starting points of the zero-order and first-order phase transitions are denoted respectively as ($T_z,\Phi_z^2$) and ($T_{c1}, \Phi_{c1}^2$). On the right side of the dashed green line in Fig.\ref{fig:grandgt}, which represents the right-most points for each line, there are no black holes present.
For a fixed value of $\Phi^2$, the system undergoes a SBH-LBH-SBH phase transition as the temperature increases, given that $\Phi_{tr}^2 < \Phi^2 < \Phi_z^2$. Additionally, for a fixed temperature of the black hole, the system also experiences a SBH-LBH-SBH phase transition as the critical value of $\Phi^2$ increases, within the range $T_z < T < T_{tr}$.

We also investigate these critical behaviors with the influences of the parameters
$\eta_1$, $\eta_2$, $\alpha$ and $\beta$, as shown in Tables.\ref{tab:gran1}-\ref{tab:gran4}.
We find that  all of the critical temperatures ($T_{c1}$, $T_{c2}$, $T_{z}$ and $ T_{tr}$) decrease and all of the critical $\Phi^2$s increase with increasing $\eta_1$.
However, all of the critical temperatures  increase and all of the critical $\Phi^2$s decrease with increasing $\eta_2$, $\alpha$ and $\beta$.

\begin{table}[!htb]
	\centering
	\begin{tabular}{|c|c|c|c|c|c|c|c|c|c|}
		\hline
		$\eta_1$ & $(\Phi^2_{c1}, T_{c1}$)&$(\Phi^2_{c2}, T_{c2}$)&$(\Phi^2_{z}, T_{z}$)&$(\Phi^2_{tr}, T_{tr}$)\\
		\hline
		0.9  & (0.828078, 0.356703) &(0.236356, 0.422563)&(0.748531,
  0.402621)&(0.522954, 0.410699)
  \\ \hline
1 &(0.833637, 0.353597)&(0.482191, 0.397275)& (0.774902,
  0.383129)& (0.651802, 0.388623)
  \\ \hline
1.1 &(0.839832, 0.350298)& (0.642647, 0.377306)&(0.799376,
  0.36794)&(0.736704, 0.371418)
  \\  \hline
	\end{tabular}
	\caption{Critical points with $\eta_1$	for $\eta_2=0.6$, $\alpha=0.4$, and $\beta=1.6$ }
	\label{tab:gran1}
\end{table}

\begin{table}[!htb]
	\centering
	\begin{tabular}{|c|c|c|c|c|c|c|c|c|c|}
		\hline
		$\eta_2$ & $(\Phi^2_{c1}, T_{c1}$)&$(\Phi^2_{c2}, T_{c2}$)&$(\Phi^2_{z}, T_{z}$)&$(\Phi^2_{tr}, T_{tr}$)\\
		\hline
		0.45& (0.883492, 0.305241)&( 0.791556, 0.317393)&( 0.85975,
  0.312801)&(0.834331, 0.314406)\\ \hline

0.55 &(0.848354, 0.337992)&(0.597715, 0.369775)&(0.800646,
  0.359017)&(0.718003, 0.363077)\\ \hline

0.65& (0.82029, 0.368788)&(0.354553, 0.42554)&(0.751269,
  0.407869)&(0.579794, 0.414905)\\
		\hline
	\end{tabular}
	\caption{Critical points with $\eta_2$	for $\eta_1=1$, $\alpha=0.4$, and $\beta=1.6$ }
	\label{tab:gran2}
\end{table}

\begin{table}[!htb]
	\centering
	\begin{tabular}{|c|c|c|c|c|c|c|c|c|c|}
		\hline
		$\alpha$ & $(\Phi^2_{c1}, T_{c1}$)&$(\Phi^2_{c2}, T_{c2}$)&$(\Phi^2_{z}, T_{z}$)&$(\Phi^2_{tr}, T_{tr}$)\\
		\hline
		0.395& (0.865026, 0.325119)&(0.73737, 0.342417)&(0.835141,
  0.336136)&(0.797332, 0.338403)\\
		\hline
0.4 &(0.848354, 0.337992)&(0.597715, 0.369775)&(0.800646,
  0.359017)&(0.718003, 0.363077)\\
		\hline
0.405& (0.832532, 0.351106)&(0.407127, 0.401536)&(0.765905,
  0.385471)&(0.613889, 0.391749)\\
		\hline
	\end{tabular}
	\caption{Critical points with $\alpha$	for $\eta_1=1$, $\eta_2=0.55$, and $\beta=1.6$}
	\label{tab:gran3}
\end{table}

\begin{table}[!htb]
	\centering
	\begin{tabular}{|c|c|c|c|c|c|c|c|c|c|}
		\hline
		$\beta$ & $(\Phi^2_{c1}, T_{c1}$)&$(\Phi^2_{c2}, T_{c2}$)&$(\Phi^2_{z}, T_{z}$)&$(\Phi^2_{tr}, T_{tr}$)\\
		\hline
		1.57&(0.867147, 0.328931)&(0.752882, 0.344235)&(0.839436,
  0.33861)&(0.806397, 0.340616)
 \\		\hline
1.61& (0.842724, 0.340924)&(0.53447, 0.379301)&(0.788289,
  0.366616)&(0.683107, 0.371478)
 \\		\hline
1.65& (0.82229, 0.352527)&(0.210372, 0.423208)&(0.741026,
  0.401698)&(0.508086, 0.410386)
  \\		\hline
	\end{tabular}
	\caption{Critical points with $\beta$	for $\eta_1=1$, $\eta_2=0.55$, and $\alpha=0.4$}
	\label{tab:gran4}
\end{table}

\subsection{ Critical behaviors of black holes in the canonical ensemble}

In the canonical ensemble with a fixed electric charge $Q$, the Helmholtz free energy is written as
\begin{eqnarray}\label{eq:helm:4d}
F_H&=&M-TS \nonumber\\
&=&\frac{r_h}{4}+\frac{\left(\alpha ^2-1\right) \left(\alpha ^2+1\right) \Lambda  r_h^{\frac{3-\alpha ^2}{\alpha ^2+1}}}{4 \left(\alpha ^2-3\right)}+\frac{\left(-\alpha ^4+2 \alpha ^2+3\right) Q^2 r_h^{\frac{\alpha ^2-1}{\alpha ^2+1}}}{2 \left(\alpha ^4-3 \alpha ^2+2\right)}\\
& &+\frac{\left(\alpha ^2+1\right) \eta _2 \left(\alpha ^2+2 \alpha  \beta -1\right) r_h^{\frac{\alpha ^2+2 \alpha  \beta +1}{\alpha ^2+1}}}{4 \left(\alpha ^2+\alpha  \beta -1\right) \left(\alpha ^2+2 \alpha  \beta +1\right)}+\frac{\alpha  \left(\alpha ^2+1\right) \eta _1 (\alpha +2 \beta ) r_h^{\frac{\alpha ^2+2 \alpha  \beta +2}{\alpha ^2+1}}}{4 \left(\alpha ^2+2 \alpha  \beta +2\right) \left(2 \alpha ^2+2 \alpha  \beta -1\right)}.\nonumber
\end{eqnarray}

The temperature of black hole  is
\begin{eqnarray}\label{eq:temp:can}
T&=&\frac{\left(\alpha ^2+1\right) Q^2 r_h^{\frac{\alpha ^2-3}{\alpha ^2+1}}}{2 \pi  \left(\alpha ^2-2\right)}-\frac{\left(\alpha ^2+1\right) \Lambda  r_h^{\frac{1-\alpha ^2}{\alpha ^2+1}}}{4 \pi }-\frac{\left(\alpha ^2+1\right) r_h^{\frac{\alpha ^2-1}{\alpha ^2+1}}}{4 \pi  \left(\alpha ^2-1\right)}\\
& &-\frac{\left(\alpha ^2+1\right) \eta _1 r_h^{\frac{\alpha  (\alpha +2 \beta )}{\alpha ^2+1}}}{4 \pi  \left(2 \alpha ^2+2 \alpha  \beta -1\right)}-\frac{\left(\alpha ^2+1\right) \eta _2 r_h^{\frac{2 \alpha  (\alpha +\beta )}{\alpha ^2+1}-1}}{4 \pi  \left(\alpha ^2+\alpha  \beta -1\right)}
,\nonumber
\end{eqnarray}
and the corresponding heat capacity is
\begin{eqnarray}
C_Q=\frac{8 \pi ^2 T r_h^{\frac{3-\alpha ^2}{\alpha ^2+1}}(\alpha ^2+1)^{-1}}{
-\frac{2 \left(\alpha ^2-3\right) Q^2 r_h^{-\frac{2}{\alpha ^2+1}}}{\alpha ^2-2}+\frac{\eta _2 \left(\alpha ^2+2 \alpha  \beta -1\right) r_h^{\frac{2 \alpha  \beta }{\alpha ^2+1}}}{\alpha ^2+\alpha  \beta -1}+\frac{\alpha  \eta _1 (\alpha +2 \beta ) r_h^{\frac{2 \alpha  \beta +1}{\alpha ^2+1}}}{2 \alpha ^2+2 \alpha  \beta -1}-\left(\alpha ^2-1\right) \Lambda  r_h^{\frac{2-2 \alpha ^2}{\alpha ^2+1}}+1}.
\end{eqnarray}

According to Eq.(\ref{eq:helm:4d}), similar to the analysis in the grand canonical system, the term $\eta_1$ vanishes
for $\alpha\rightarrow 0$, which also means the term $\eta_2$ is significantly more important in the case without  dilaton field.
However, in the case coupling with the dilaton field, i.e.$\alpha\neq 0$, considering the asymptotic behaviors of the Helmholtz free energy, the term $\eta_1$ plays a more important role than $\eta_2$ as $r_h\rightarrow \infty$.

Using the condition of inflection point in VdW system
\begin{equation}
\frac{\partial T}{\partial r_h}\left|_{Q=Q_c,r_h=r_c}\right.
=\frac{\partial^2 T}{\partial r_h^2}\left|_{Q={Q_c},r_h=r_c}\right.
=0,
\end{equation}

For given a fixed $Q$, we can receive the critical $Q_{c}$ and the equation of critical radius of black hole as
\begin{align}
Q_{c}^2=&\frac{\left(\alpha ^2-2\right) r_c^{\frac{2}{\alpha ^2+1}}}{2 \left(\alpha ^2-3\right)}
-\frac{\left(\alpha ^4-3 \alpha ^2+2\right) \Lambda  r_c^{\frac{4-2 \alpha ^2}{\alpha ^2+1}}}{2 \left(\alpha ^2-3\right)}
+\frac{\left(\alpha ^2-2\right) \eta _2 \left(\alpha ^2+2 \alpha  \beta -1\right) r_c^{\frac{2 \alpha  \beta +2}{\alpha ^2+1}}}{2 \left(\alpha ^2-3\right) \left(\alpha ^2+\alpha  \beta -1\right)}\nonumber\\
&+\frac{\alpha  \left(\alpha ^2-2\right) \eta _1 (\alpha +2 \beta ) r_c^{\frac{2 \alpha  \beta +3}{\alpha ^2+1}}}{2 \left(\alpha ^2-3\right) \left(2 \alpha ^2+2 \alpha  \beta -1\right)},
\label{eq:qcri}\\
0=&\frac{\eta _2 (\alpha  \beta +1) \left(\alpha ^2+2 \alpha  \beta -1\right) r_c^{\frac{2 \alpha  \beta }{\alpha ^2+1}}}{\alpha ^2+\alpha  \beta -1}+\frac{\alpha  \eta _1 (\alpha +2 \beta ) (2 \alpha  \beta +3) r_c^{\frac{2 \alpha  \beta +1}{\alpha ^2+1}}}{4 \alpha ^2+4 \alpha  \beta -2}+\left(\alpha ^4-3 \alpha ^2+2\right) \Lambda  r_c^{\frac{2-2\alpha^2}{\alpha ^2+1}}+1\label{eq:qcradi}.
\end{align}

\begin{figure}[H]
	 \subfigure[  ]{\label{fig:alpha0:1} 
  \includegraphics[width=8cm]{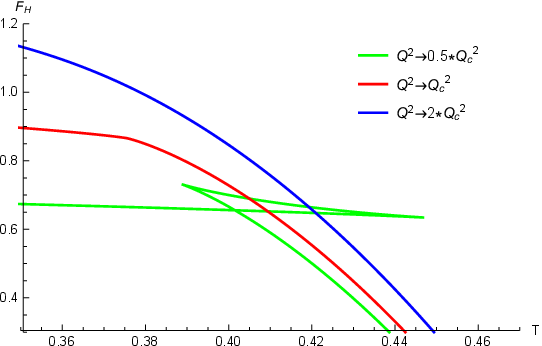}}%
  \subfigure[ ]{\label{fig:alpha0:2} 
  \includegraphics[width=8cm]{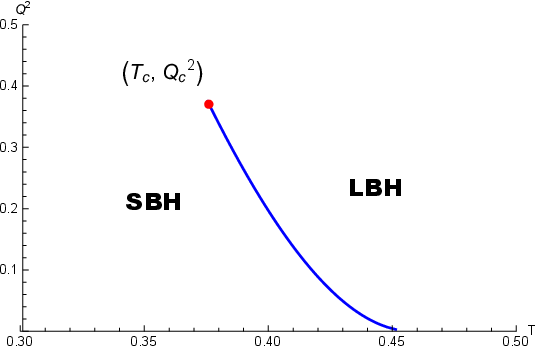}}%
	\caption{Critical behaviors with $Q^2$ for $\alpha=0$, $\eta_1=0.1$, and $\eta_2=2$}	
	\label{fig:alpha0}
\end{figure}

When $\alpha=0$ in the AdS spacetime, from Eq.\eqref{eq:qcradi},
we can find it has only one positive root
$r_c=\frac{\sqrt{{\eta_2}+1}}{2}$,
which depends only on $\eta_2$.
Then, the critical values of  $Q^2$ and  the corresponding temperature are respectively calculated as
\begin{align}\label{eq:cri values0}
 Q^2_{c}=\frac{1}{24}  ({\eta_2}+1)^2,\quad
T_c=\frac{ 2\sqrt{{\eta_2}+1}}{3 \pi }+\frac{ {\eta_1}}{4 \pi },
\end{align}
which suggest there could be first-order SBH/LBH phase transitions in this case as shown in Fig.\ref{fig:alpha0}.

\begin{figure}[H]
	 \subfigure[  ]{\label{fig:cgt:1} 
  \includegraphics[width=6cm]{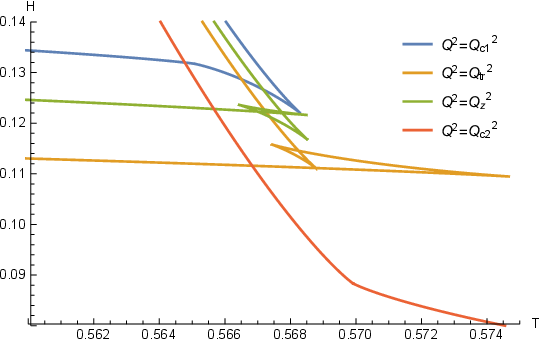}}%
  \subfigure[ ]{\label{figc:cgt:2} 
  \includegraphics[width=6cm]{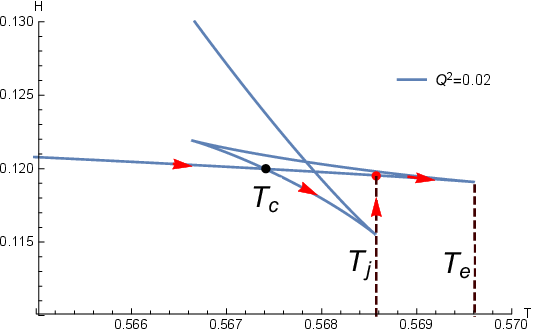}}%
  \subfigure[ ]{\label{figc:cgt:3} 
  \includegraphics[width=6cm]{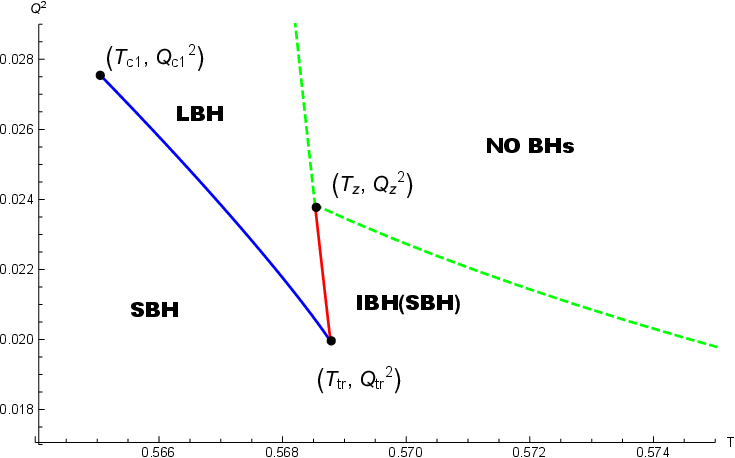}}
	\caption{Critical behaviors for $\eta_1= 0.58$, $\eta_2=0.41$, $\alpha =0.6$, and $\beta =0.8$}	
	\label{figc:cgt}
\end{figure}

When $ \alpha \neq 0 $,
similar to that in the grand canonical ensemble when studying  the critical behaviors,
there are several conditions to be considered,
i.e.
all of the roots of Eq.\eqref{eq:qcradi}, the $Q^2_c$ of Eq.\eqref{eq:qcri}, and the temperatures of black holes corresponding to the critical points must  be positive.
Based on these, through numerical analysis,  there  are at most two critical radii.
Taking the case of $ \alpha = 0.6$, and $\beta = 0.8$ as an example, there can be at most two critical radii  for $\eta_1 >0 $. 
By investigating the Helmholtz free energy, as shown in Fig.\ref{figc:cgt},
we find that  there are also reverse reentrant phase transitions among SBHs, LBHs and SBHs.
We also investigate the influences of the parameters
$\eta_1$, $\eta_2$, $\alpha$ and $\beta$ on these critical behaviors as shown in Tables.\ref{tab:can1}-\ref{tab:can4}.
We find that all of the critical temperatures ($T_{c1}$, $T_{c2}$, $T_{z}$ and $ T_{tr}$) decrease and all of the critical charges ($Q^2_{c1}$, $Q^2_{c2}$, $Q^2_{z}$ and $ Q^2_{tr}$) increase with increasing $\eta_1$.
However, all of the critical temperatures  increase and all of the critical charges ($Q^2$s) decrease with increasing $\eta_2$, $\alpha$ and $\beta$.

\begin{table}[!htb]
	\centering
	\begin{tabular}{|c|c|c|c|c|c|c|c|c|c|}
		\hline
		$\eta_1$ & $(Q^2_{c1}, T_{c1}$)&$(Q^2_{c2}, T_{c2}$)&$(Q^2_{z}, T_{z}$)&$(Q^2_{tr}, T_{tr}$)\\
		\hline
		0.58 &(0.0275472, 0.565051)&(0.00461282, 0.569917)&(0.0238421,
  0.568524)&(0.0199481, 0.568775)
 \\		\hline
0.59&(0.0282922, 0.563862)&(0.013105, 0.567242)&(0.0254134,
  0.566206)&(0.0227447, 0.566408)
 \\		\hline
0.6&(0.0291293, 0.562626)&(0.0200776, 0.564737)&(0.0270936,
  0.564038)&(0.0254712, 0.564185)
  \\		\hline
	\end{tabular}
	\caption{The critical points with $\eta_1$	for $\eta_2=0.41$, $\alpha=0.6$, and $\beta=0.8$}
	\label{tab:can1}
\end{table}

\begin{table}[!htb]
	\centering
	\begin{tabular}{|c|c|c|c|c|c|c|c|c|c|}
		\hline
		$\eta_2$ & $(Q^2_{c1}, T_{c1}$)&$(Q^2_{c2}, T_{c2}$)&$(Q^2_{z}, T_{z}$)&$(Q^2_{tr}, T_{tr}$)\\
		\hline
		0.4 &(0.0297303, 0.557388)&(0.0201646, 0.559553)&(0.0276001,
  0.558839)&(0.0258868, 0.558989)
\\		\hline
0.41&(0.0282922, 0.563862)&(0.013105, 0.567242)&(0.0254134,
  0.566206)&(0.0227447, 0.566408)
\\		\hline
0.42&(0.0270065, 0.570258)&(0.00524945, 0.575021)&(0.0234473,
  0.57365)&(0.0197409, 0.573899)
  \\		\hline
	\end{tabular}
	\caption{The critical points with $\eta_2$	for $\eta_1=0.59$, $\alpha=0.6$, and $\beta=0.8$}
	\label{tab:can2}
\end{table}

\begin{table}[!htb]
	\centering
	\begin{tabular}{|c|c|c|c|c|c|c|c|c|c|}
		\hline
		$\alpha$ & $(Q^2_{c1}, T_{c1}$)&$(Q^2_{c2}, T_{c2}$)&$(Q^2_{z}, T_{z}$)&$(Q^2_{tr}, T_{tr}$)\\
		\hline
		0.599 &(0.029664, 0.559358)&(0.0198809, 0.561567)&(0.0275017,
  0.560842)&(0.0257504, 0.560993)
\\		\hline
0.6&(0.0282922, 0.563862)&(0.013105, 0.567242)&(0.0254134,
  0.566206)&(0.0227447, 0.566408)
\\		\hline
0.601&(0.0270375, 0.568381)&(0.00544345, 0.573117)&(0.0234916,
  0.571751)&(0.0198093, 0.571999)
  \\		\hline
	\end{tabular}
	\caption{The critical points with $\alpha$ for $\eta_1=0.59$, $\eta_2= 0.41$, and $\beta=0.8$}
	\label{tab:can3}
\end{table}

\begin{table}[!htb]
	\centering
	\begin{tabular}{|c|c|c|c|c|c|c|c|c|c|}
		\hline
		$\beta$ & $(Q^2_{c1}, T_{c1}$)&$(Q^2_{c2}, T_{c2}$)&$(Q^2_{z}, T_{z}$)&$(Q^2_{tr}, T_{tr}$)\\
		\hline
		0.797 &(0.0297257, 0.5606)&(0.0205733, 0.562678)&(0.0276622,
  0.56199)&(0.0260213, 0.562135)
\\		\hline
0.801&(0.0278492, 0.564943)&(0.0103712, 0.568806)&(0.0247118,
  0.567649)&(0.0216714, 0.567869)
\\		\hline
0.804&(0.0266061, 0.568168)&(0.00141926, 0.573625)&(0.0227377,
  0.572095)&(0.0185209, 0.572364)
  \\		\hline
	\end{tabular}
	\caption{The critical points with $\beta$for $\eta_1=0.59$, $\eta_2= 0.41$,and $\alpha=0.6$}
	\label{tab:can4}
\end{table}

\section{Conclution}\label{4s}

In  this paper,  we discuss Maxwell-dilaton  massive
gravity including a nonminimal coupling  term  between the graviton and the dilaton field and obtain charged solutions of a dilatonic black hole in four dimensional spacetime. Here the dilaton potential $V(\varphi)$ takes a  Liouville-type form, where the last two terms of the potential are associated with the graviton terms.

we discuss the singularity of the solution from the Kretschmann scalar and find that the horizons of  black holes are simply singularities of coordinates,  whereas there is an essential singularity located  at
the origin. Moreover, the asymptotic behavior of the solutions is neither asymptotically flat nor  asymptotically (A)dS. We show that the black hole solutions can provide one horizon, two horizons, and extreme and naked singularity black holes for suitably fixed parameters. To verify the first law of black hole thermodynamics, we calculate the thermodynamic  quantities of these dilatonic  black
holes, such as the entropy, temperature, and mass. Then, we  pay  further  attention  to  phase  space  and  find  richer and more interesting critical phenomena the in grand canonical and canonical ensembles.

In grand canonical ensembles, 
there are no VdW-like critical behaviors of black holes  without the coupling effect between the graviton and the dilaton field (i.e.$\gamma = 0$). 
However, there are more critical phenomena  when $\alpha\neq 0$.
There may be one-order VdW-like phase transitions between SBHs and LBHs.
More interestingly, by investigating the Gibbs free energy,
there may also be new and interesting critical behaviors with a triple critical point and two types of phase transitions (zero and one-order),
known as zero-order reverse reentrant phase transition
among SBHs, LBHs and SBHs,
because its thermodynamic process is the reverse of that of reentrant phase transition owing to the minimum of the global Gibbs free energy. In the canonical ensembles,
there are only VdW-like phase transitions in the phase space of black holes without the coupling. However, by
investigating the Helmholtz free energy,
we find that there also exist the reverse reentrant phase transitions among SBHs, LBHs and SBHs when $\alpha\neq 0$.

{\bf Acknowledgements}

This work is supported  by National Key
Research and Development Program of China under Grant
No. 2020YFC2201400.  D. C. Zou is supported by National Natural Science Foundation of China (Grant No. 12365009) and Natural Science Foundation of Jiangxi Province (No. 20232BAB201039).


\begin{thebibliography}{}


\bibitem{Riess:1998}
A. G.Riess, \textsl{et\ al.}(Supernova Search Team),
Astron.\ J\ {\bf116}, 1009(1998);\

\bibitem{Perlmutter:1999}
Perlmutter, \textsl{ et\ al.}(Surpernova Cosmoly Project),
Astrophys.\ J\ {\bf517}, 565(1999).

\bibitem{Planck:2015fie}
P.~A.~R.~Ade \textit{et al.} [Planck],
Astron. Astrophys. \textbf{594}, A13 (2016)
[arXiv:1502.01589 [astro-ph.CO]];\

\bibitem{WMAP:2003elm}
D.~N.~Spergel \textit{et al.} [WMAP],
Astrophys. J. Suppl. \textbf{148}, 175-194 (2003)
[arXiv:astro-ph/0302209 [astro-ph]].

\bibitem{Weinberg:1965pg}
S. Weinberg,
Phys.\ Rev.\ B\ {\bf138},988(1965);\
\bibitem{Boulware:1975cg}
D. G.Boulware, and S. Deser,
Annals\ of\ Physics\ {\bf89}, 193(1975).

\bibitem{Fierz:1939}
M.Fierz, and W.Pauli,
Proc.\ R.\ Soc.\ Lond.\ A\ {\bf173}, 211(1939).

\bibitem{Boulware:1972if}
D. G.Boulware, and S.Desser,
Phys.\ Lett.\ B\ {\bf40}, 227(1972);\ 
\bibitem{Boulware:1972fr}
D. G.Boulware, and S.Deser,
Phys.\ Rev.\ D\ {\bf6}, 3368(1972).
\bibitem{Rham:2011tl}
C.de Rham, G.Gabadadze, and A. J.Tolley,
Phys.\ Rev.\ Lett.\ {\bf106}, 231101(2011);\
\bibitem{Rham:2014mg}
C.de Rham,
Living\ Rev.\ Relativity\ {\bf17}, 7(2014).
\bibitem{Hinterbichler:2012}
K.~Hinterbichler,
Rev.\ Mod.\ Phys.\  {\bf 84}, 671 (2012);\
\bibitem{Vegh:2013}
D. Vegh,
arXiv:\ 1301.0537\ [hep-th];\


\bibitem{Cai:2015tb}
R. G.Cai,\ Y. P.Hu,\ Q. Y.Pan,\ and Y. L.Zhang,\
Phys.\ Rev.\ D\ {\bf91}, 024032(2015)
[arXiv:1409.2369 [hep-th]].

\bibitem{Xu:2015rfa}
J.~Xu, L.~M.~Cao and Y.~P.~Hu,
Phys. Rev. D \textbf{91} (2015) no.12, 124033
[arXiv:1506.03578 [gr-qc]].

\bibitem{Zou:2016sab}
D.~C.~Zou, R.~Yue and M.~Zhang,
Eur. Phys. J. C \textbf{77} (2017) no.4, 256
[arXiv:1612.08056 [gr-qc]].

\bibitem{Zou:2017juz}
D.~C.~Zou, Y.~Liu and R.~H.~Yue,
Eur. Phys. J. C \textbf{77} (2017) no.6, 365
[arXiv:1702.08118 [gr-qc]].


\bibitem{Babichev:2014rb}
E.Babichev,\ and A.Fabbri,\
Phys.\ Rev.\ D\ {\bf90}, 084019 (2014);\
\bibitem{Hendi:2015eb}
S. H. Hendi,\ B. E.Panah,\ and S.Panahiyan,\
J\ High\ Energy\ Phys.\ {\bf2015}, 157(2015);\
\bibitem{Zhang:2017lhl}
M.~Zhang, D.~C.~Zou and R.~H.~Yue,
Adv. High Energy Phys. \textbf{2017} (2017), 3819246
[arXiv:1707.04101 [hep-th]].
A.Ace\~{n}a,\ E.L\'{o}pez,\ and M.Llerena,\
Phys.\ Rev.\ D\ {\bf97}, 064043(2018).

\bibitem{Kodama:2014ss}
H.Kodama,\ and I.Arraut,\
Progress\ of\ Theoretical\ and\ Experimental\ Physics\ {\bf2014}, 023E02(2014).

\bibitem{Hendi:2016jx}
S. H.Hendi,\ G. Q.Li,\ J. X.Mo,\ S.Panahiyan,\ and B. E.Panah,\
Euro.\ Phys.\ J\  C\ {\bf76}, 571(2016);\
\bibitem{Hendi:2018td}
S. H.Hendi,\ and M.Momennia,\
arXiv:\ 1801.07906;\
\bibitem{Hendi:2018gb}
S. H.Hendi,\ B.Eslam Panah,\ and S.Panahiyan,\
Fortschritte\ der\ Physik\ {\bf66}, 1800005(2018).




\bibitem{Green:1987sw}
M.~B.~Green, J.~H.~Schwarz and E.~Witten,
Superstring Theory, (Cambridge University Press, Cambridge 1987).

\bibitem{Mignemi:1991wa}
S.~Mignemi and D.~L.~Wiltshire,
Phys. Rev. D \textbf{46}, 1475-1506 (1992)
[arXiv:hep-th/9202031 [hep-th]];\
\bibitem{Poletti:1994ff}
S.~J.~Poletti and D.~L.~Wiltshire,
Phys. Rev. D \textbf{50}, 7260-7270 (1994)
[erratum: Phys. Rev. D \textbf{52}, 3753-3754 (1995)]
[arXiv:gr-qc/9407021 [gr-qc]];\
\bibitem{Poletti:1994ww}
S.~J.~Poletti, J.~Twamley and D.~L.~Wiltshire,
Phys. Rev. D \textbf{51}, 5720-5724 (1995)
[arXiv:hep-th/9412076 [hep-th]];\
\bibitem{Cai:1997ii}
R.~G.~Cai, J.~Y.~Ji and K.~S.~Soh,
Phys. Rev. D \textbf{57}, 6547-6550 (1998)
[arXiv:gr-qc/9708063 [gr-qc]];\
\bibitem{Clement:2002mb}
G.~Clement, D.~Gal'tsov and C.~Leygnac,
Phys. Rev. D \textbf{67}, 024012 (2003)
[arXiv:hep-th/0208225 [hep-th]].

\bibitem{Dehghani:2004sa}
M.~H.~Dehghani and N.~Farhangkhah,
Phys. Rev. D \textbf{71}, 044008 (2005)
[arXiv:hep-th/0412049 [hep-th]].

\bibitem{Sheykhi:2007wg}
A.~Sheykhi,
Phys. Rev. D \textbf{76}, 124025 (2007)
[arXiv:0709.3619 [hep-th]].
\bibitem{Gao:2004tu}
C.~J.~Gao and S.~N.~Zhang,
Phys. Rev. D \textbf{70}, 124019 (2004)
[arXiv:hep-th/0411104 [hep-th]];\
\bibitem{Gao:2004tv}
C.~J.~Gao and S.~N.~Zhang,
Phys. Lett. B \textbf{605}, 185-189 (2005)
[arXiv:hep-th/0411105 [hep-th]];\

\bibitem{KordZangeneh:2015fdy}
M.~Kord Zangeneh, M.~H.~Dehghani and A.~Sheykhi,
Phys. Rev. D \textbf{92}, no.10, 104035 (2015)
doi:10.1103/PhysRevD.92.104035
[arXiv:1509.05990 [gr-qc]].

\bibitem{KordZangeneh:2015hfy}
M.~Kord Zangeneh, A.~Sheykhi and M.~H.~Dehghani,
Phys. Rev. D \textbf{91}, no.4, 044035 (2015)
[arXiv:1505.01103 [gr-qc]].

\bibitem{Dehyadegari:2017flm}
A.~Dehyadegari, A.~Sheykhi and A.~Montakhab,
Phys. Rev. D \textbf{96}, no.8, 084012 (2017)
[arXiv:1707.05307 [hep-th]].



\bibitem{Berti:2015itd}
E.~Berti, E.~Barausse, V.~Cardoso, L.~Gualtieri, P.~Pani, U.~Sperhake, L.~C.~Stein, N.~Wex, K.~Yagi and T.~Baker, \textit{et al.}
Class. Quant. Grav. \textbf{32}, 243001 (2015)
[arXiv:1501.07274 [gr-qc]].

\bibitem{Pani:2011xm}
P.~Pani, E.~Berti, V.~Cardoso and J.~Read,
Phys. Rev. D \textbf{84}, 104035 (2011)
[arXiv:1109.0928 [gr-qc]].
\bibitem{Yunes:2011we}
N.~Yunes and L.~C.~Stein,
Phys. Rev. D \textbf{83}, 104002 (2011)
[arXiv:1101.2921 [gr-qc]].
\bibitem{Kanti:1995vq}
P.~Kanti, N.~E.~Mavromatos, J.~Rizos, K.~Tamvakis and E.~Winstanley,
Phys. Rev. D \textbf{54}, 5049-5058 (1996)
[arXiv:hep-th/9511071 [hep-th]].
\bibitem{Torii:1996yi}
T.~Torii, H.~Yajima and K.~i.~Maeda,
Phys. Rev. D \textbf{55}, 739-753 (1997)
[arXiv:gr-qc/9606034 [gr-qc]].
\bibitem{Guo:2008hf}
Z.~K.~Guo, N.~Ohta and T.~Torii,
Prog. Theor. Phys. \textbf{120}, 581-607 (2008)
[arXiv:0806.2481 [gr-qc]].
\bibitem{Ohta:2009tb}
N.~Ohta and T.~Torii,
Prog. Theor. Phys. \textbf{121}, 959-981 (2009)
[arXiv:0902.4072 [hep-th]].
\bibitem{Ohta:2009pe}
N.~Ohta and T.~Torii,
Prog. Theor. Phys. \textbf{122}, 1477-1500 (2009)
[arXiv:0908.3918 [hep-th]].

\bibitem{Kleihaus:2011tg}
B.~Kleihaus, J.~Kunz and E.~Radu,
Phys. Rev. Lett. \textbf{106}, 151104 (2011)
[arXiv:1101.2868 [gr-qc]].

\bibitem{Maselli:2015tta}
A.~Maselli, P.~Pani, L.~Gualtieri and V.~Ferrari,
Phys. Rev. D \textbf{92}, no.8, 083014 (2015)
[arXiv:1507.00680 [gr-qc]].



\bibitem{Liu:2023sxw}
B.~Liu, R.~H.~Yue, D.~C.~Zou, L.~Zhang, Z.~Y.~Yang and Q.~Pan,
[arXiv:2310.14653 [gr-qc]].


\bibitem{Wald:1993nt}
R.~M.~Wald,
Phys. Rev. D \textbf{48}, no.8, R3427-R3431 (1993)
[arXiv:gr-qc/9307038 [gr-qc]].

\bibitem{Iyer:1994ys}
V.~Iyer and R.~M.~Wald,
Phys. Rev. D \textbf{50}, 846-864 (1994)
[arXiv:gr-qc/9403028 [gr-qc]].

\bibitem{Abbott:1982}
L.F.Abbott and S.Deser,\ \
Nucl.\ phys.\ \textbf{B195},\ 76 (1982)






\end{thebibliography}
\end{document}